\documentclass[%
twocolumn, superscriptaddress,
nofootinbib,
 amsmath,amssymb,
 aps,
pra,
]{revtex4-1}
\usepackage[all]{xy}
\usepackage{mathrsfs}
\usepackage{graphicx}
\usepackage{epstopdf}
\usepackage{dcolumn}
\usepackage{bm}
\usepackage{soul}
\usepackage{color} 
\usepackage{CJK}
\usepackage{ctable}  
\usepackage{hyperref}
\usepackage{dsfont}

\newcommand{\beq}{\begin{equation}}
\newcommand{\eeq}{\end{equation}}
\newcommand{\beqa}{\begin{eqnarray}}
\newcommand{\eeqa}{\end{eqnarray}}

\newcommand{\ua}{\uparrow}
\newcommand{\da}{\downarrow}

\definecolor{myorange}{RGB}{255,117,40}

\DeclareMathOperator{\tr}{tr}

\begin{document}
\title{Thermalization with detailed-balanced two-site Lindblad dissipators}
\author{Mikel Palmero}
\email{palmerolazcoz@sutd.edu.sg}
\affiliation{Science and Math Cluster, Singapore University of Technology and Design, 8 Somapah Road, 487372 Singapore}
\author{Xiansong Xu}
\affiliation{Science and Math Cluster, Singapore University of Technology and Design, 8 Somapah Road, 487372 Singapore}
\author{Chu Guo}
\affiliation{Zhengzhou Information Science and Technology Institute, Zhengzhou 450004, China}

\author{Dario Poletti}
\email{dario\_poletti@sutd.edu.sg} 
\affiliation{Science and Math Cluster, Singapore University of Technology and Design, 8 Somapah Road, 487372 Singapore}
\affiliation{Engineering Product Development, Singapore University of Technology and Design, 8 Somapah Road, 487372 Singapore}
\begin{abstract}
  The use of two-site Lindblad dissipator to generate thermal states and study heat transport raised to prominence since [J. Stat. Mech. (2009) P02035] by Prosen and \v{Z}nidari\v{c}. Here we propose a variant of this method based on detailed balance of internal levels of the two site Hamiltonian and characterize its performance. We study the thermalization profile in the chain, the effective temperatures achieved by different single and two-site observables, and we also investigate the decay of two-time correlations. We find that at a large enough temperature the steady state approaches closely a thermal state, with a relative error below $1\%$ for the inverse temperature estimated from different observables.    
\end{abstract}
\maketitle
\section{Introduction}

The ability to control heat flow at the nanoscale can offer important technological opportunities. In many-body quantum systems the interactions can significantly change the transport properties, e.g., turning it from ballistic to diffusive \cite{Prosen2011}. Recently it was also shown that interactions can be instrumental in realizing highly performing diodes \cite{BalachandranPoletti2018, BalachandranPoletti2018a}. Of particular importance is the current between heat baths at different temperatures, i.e., heat current. However, the study of transport in the presence of heat baths requires an appropriate modelling of the baths which can be extended to intermediate-size and large many-body quantum systems. 

Different approaches have been used to model the coupling of a many-body system to a thermal bath, some global and some local. In global approaches the model considers the Hamiltonian of the entire system, and it usually requires to diagonalize it, which is something that can be done only for limited system sizes. One of the global approaches uses master equations in Gorini-Kossakovski-Sudarshan-Lindblad (GKSL) form \cite{GoriniSudarshan1976, Lindblad1976} (see also \cite{BreuerPetruccione2007}). The use of this method, however, can result in an unphysical description of the current within a system. In fact, it can predict that a heat current flows from one bath to the system, and from the system to the other bath, while no heat current flows within the system \cite{WichterichMichel2007}. 
Another global approach that gives more accurate results for the heat current within the chain is the Redfield master equation \cite{Redfield1957}. Recently it was even shown how to implement a Redfield master equation on a large system with matrix product states and operators \cite{XuPoletti2019}, although only for short-time evolutions. Other approaches to study many-body quantum systems in contact with a finite temperature bath of harmonic oscillators are based on the star-to-chain transformation, which converts the bath to a bosonic chain \cite{ChinPlenio2010,PriorPlenio2010, deVegaBanuls2015, GuoPoletti2018, CasciodeVega2018}. This strategy is particularly effective if preceded by a thermofield transformation which maps a finite temperature bath to two zero-temperature baths \cite{deVegaBanuls2015, GuoPoletti2018, CasciodeVega2018}. As of now, this approach is limited to relatively short-time evolutions and to noninteracting baths. For the cases in which it is possible to assume translational invariance in the transfer tensors describing the time evolution, it is possible to reach longer times \cite{CerrilloCao2014, RosenbachPlenio2016}.         

A few approaches have been proposed for the long-time, or even steady state, study of many-body quantum systems in contact with a heat bath \cite{GelmanKosloff2004, KatzKosloff2008, TorronteguiKosloff2016, ReichentalPodolsky2018, ProsenZnidaric2009}. Further investigation is needed to clarify which of these methods performs better, depending on the properties of both the systems and the baths studied. 
Here we focus on the two-site thermalization approach proposed in \cite{ProsenZnidaric2009}, which is readily implemented with matrix product states algorithms, and which has already been used to investigate transport physics \cite{ZnidaricRossini2010, Znidaric2011, Znidaric2011a, Mendoza-ArenasJaksch2015}. In this case, the dissipative part of the master equation in GKSL form acts simultaneously on the two sites closest to each edge trying to thermalize them. This approach improves significantly the results for local master equations whose dissipator acts on a single site.     Here, we propose a modification of the scheme presented in \cite{ProsenZnidaric2009} to obtain a modified two-site dissipator and study how it thermalizes the system of interest. 

This paper is divided in the following sections: In Sec.~\ref{sec:model} we describe the spin chain we aim to thermalize, and in Sec.~\ref{sec:method} we describe our approach for two-site thermalizing GKSL master equations. In Sec.~\ref{sec:results} we characterize the performance of our method both with static and dynamical observables, and in Sec.~\ref{sec:conclusions} we draw our conclusions. 

\section{Spin chain Hamiltonian}\label{sec:model}     

For the two-site dissipator to be able to thermalize the system, as we will see in more detail later, we need to consider a non-integrable Hamiltonian. We thus consider the following, one-dimensional spin Hamiltonian 
\begin{equation}
\label{XXZHamiltonian}
H = \sum_{l=1}^{L} ( h_x\sigma_l^x+h_z\sigma_l^z ) 
+ \sum_{l=1}^{L-1} J(\sigma_l^x\sigma_{l+1}^x+\sigma_l^y\sigma_{l+1}^y)
+\Delta\sigma_l^z\sigma_{l+1}^z,
\end{equation} 
for a chain of length $L$, where $h_x$ is the magnitude of the transverse field, $h_z$ the longitudinal field, $J$ the tunneling amplitude and $\Delta$ the nearest neighbour interaction. The $\sigma_l^a$ with $a=x,\;y,\;z$ are operators associated to the Pauli matrices. This Hamiltonian is such that when $h_x=0$, or $h_x=J=0$ or $h_z=J=0$ the system is integrable. For $J=0$ we recover the Ising chain with transverse and longitudinal field, whose chaotic properties were analyzed in Ref.~\cite{Mejia-MonasterioCasati2005}. For $J\ne 0$ we get an $XXZ$ chain with a transverse and a longitudinal field.  

When the Hamiltonian is integrable, the statistics of the energy level spacing follows a Poisson distribution. On the contrary, non-integrable systems can have a level spacing statistics which follows a Wigner-Dyson distribution. Following Ref.~\cite{OganesyanHuse2007}, it is possible to characterize how closely to a Poisson or a Wigner-Dyson distribution the system is. In Ref.~\cite{OganesyanHuse2007} the authors proposed to compute the average of the ratio between all the consecutive energy differences. This means that if $\delta_n$ is the energy difference between level $n$ and $n+1$, they compute $\bar{r}$, which is the arithmetic average of the $r_n$ where $r_n=\min(\delta_n/\delta_{n+1},\delta_{n+1}/\delta_{n})$ is the ratio between two consecutive energy difference. For a Hamiltonian whose level spacing follows a Poisson distribution, $\bar{r}=2\ln2 -1 \approx 0.386$, while for the Wigner-Dyson $\bar{r}\approx 0.5295$ \cite{OganesyanHuse2007}.      
Note that for $J_x=0$ the Hamiltonian conserves the total magnetization, while for $J_x\ne 0$ the Hamiltonian has a reflection symmetry $j\rightarrow L-j$. We thus have to restrict ourselves to distinct number conserving sectors of the Hamiltonians to characterize the statistics of the energy levels.       
\section{Two-site thermalizing baths}\label{sec:method}   
In order to thermalize the spin chain described in Sec. \ref{sec:model}, we apply a dissipator $\mathcal{D}$ in Lindblad form at one or two edges of the chain. The master equation is thus written as
\begin{align}
\label{Lindbladian}     
\dot{\rho} &= -\frac{i}{\hbar}[H,\rho] + \mathcal{D}[\rho] \nonumber \\ 
&= -\frac{i}{\hbar}[H,\rho] + \sum \gamma_{p,q}\left( \Gamma_p^{}\rho \Gamma_q^\dagger -\frac{1}{2}\left\{ \Gamma_p^\dagger \Gamma_q^{}, \rho \right\}\right),
\end{align}
where $\rho$ is the density operator, $\Gamma_p^{}$ are possible jump operators which act locally on one or the other edge, $\gamma_{p,q}^{}$ the corresponding weight of each combination of jump operators and $\left\{ \cdot, \cdot\right\}$ stands for the anti-commutator. With $\mathcal{L}$ we indicate the superoperator acting on the density operator $\rho$. Note that hereafter we set $\hbar=k_B=1$, where $k_B$ is the Boltzmann constant, and the unit of energy is set such that $\Delta=-1$. The matrix formed by the elements $\gamma_{p,q}^{}$ has to be non-negative to guarantee that the density matrix $\rho$ is physical at all times \cite{GoriniSudarshan1976, Lindblad1976}. In the following we use jump operators which consists of all the possible combinations of two-site operators composed by the identity matrix and the Pauli matrices. The task is then to figure out the elements $\gamma_{p,q}^{}$ such that the overall chain approaches a thermal state in the steady state. In Ref.~\cite{ProsenZnidaric2009} the authors chose the $\gamma_{p,q}^{}$ such that if the system comprised of only two sites, the steady state would be exactly the thermal state of an appropriately chosen two-site Hamiltonian. However this condition does not uniquely define the $\gamma_{p,q}^{}$. In fact, this condition only ensures that the steady state is correct, but it does not fix the decay rate of all the other eigenmodes of Eq. (\ref{Lindbladian}), the so-called {\it rapidities}. Hence there is significant freedom in the choice of the jump operators $\Gamma_p^{}$ and dissipation rate matrix $\gamma_{p,q}^{}$.  
The choice to build the coefficients $\gamma_{p,q}^{}$, and the operators $\Gamma_p^{}$, done in Ref.~\cite{ProsenZnidaric2009} is such that all rapidities, non-corresponding to the steady state, decay quickly at the same rate. We will thus refer to this approach as the {\it fast relaxation} approach.   

In the following we detail a different approach which is insipired by Ref.~\cite{ProsenZnidaric2009}, but that uses a different condition to determine uniquely the master equation. This approach is based on maintaining detailed balance between the four energy levels of the two-site Hamiltonian which the bath is coupled to. We describe this method, which we refer to as {\it detailed balance} approach, in the following subsection.     

\subsection{Detailed balance approach}\label{ssec:our}          

The dissipator is designed to generate a thermal state on the two sites it acts upon. The two sites form a four-level system (levels $0$ to $3$ in increasing energies $E_0$ to $E_3$) and hence there are $6$ possible transitions and $12$ possible jump operators if we consider raising and lowering rates between all possible levels. In this rotated basis we can write the target density operator $\rho_T^{}$ as    
\begin{align}
\label{mixed_state}
\rho_T^{} = \sum_{j=0}^3 P_j |j\rangle\langle j|,
\end{align}
where the $P_j$ are given by 
\begin{align}
\label{mixed_state_weights}
P_j = \frac{e^{-\beta_B E_j}}{\sum_i e^{-\beta_B E_i}},
\end{align}
where the $E_i$ are the energies of the target two-site (i.e., four-level) Hamiltonian $H_{2S}$ and $\beta_B$ is the target inverse temperature that the bath will try to set. The master equation can then be written as 
\begin{align}
\label{Lindbladian_simp}
\dot{\rho} & =  -i[H_{2S},\rho] \nonumber \\ 
  & + \gamma\!\!\!\!\sum_{n,m>n} \!\left[ \lambda^+_{(n,m)}\left( \Gamma^\dagger_{(n,m)}\rho \Gamma_{(n,m)}^{} -\frac{1}{2}\left\{ \Gamma^\dagger_{(n,m)} \Gamma_{(n,m)}^{}, \rho \right\}\right) \right. \nonumber \\ 
  &+ \left. \lambda^-_{(n,m)}\left( \Gamma_{(n,m)}^{}\rho \Gamma^\dagger_{(n,m)} -\frac{1}{2}\left\{ \Gamma_{(n,m)}^{} \Gamma^\dagger_{(n,m)}, \rho \right\}\right) \right]       
\end{align}
where $\Gamma_{(n,m)}^{}=|n\rangle\langle m|$ (or $\Gamma^\dagger_{(n,m)}=|m\rangle\langle n|$) is the operator that lowers (raises) the four-level system from level $m$ to $n$ ($n$ to $m$), $\gamma$ is a global relaxation rate, and $\lambda^+_{(n,m)}$ and $\lambda^-_{(n,m)}$ are respectively the relative rate of raising state $n$ to $m$ or lowering $m$ to $n$.  
Using Eq. (\ref{mixed_state}) in Eq. (\ref{Lindbladian_simp}), and imposing the steady-state condition (i.e., $d\rho/dt = 0$), we obtain the following set of equations    
\begin{align}
0 =& P_1\lambda_{(0,1)}^- - P_0\lambda_{(0,1)}^+ + P_2\lambda_{(0,2)}^- -P_0\lambda_{(0,2)}^+ \nonumber\\ 
&+P_3\lambda_{(0,3)}^- - P_0\lambda_{(0,3)}^+\nonumber\\
0 =& -P_1\lambda_{(0,1)}^- + P_0\lambda_{(0,1)}^+ + P_2\lambda_{(1,2)}^- -P_1\lambda_{(1,2)}^+ \nonumber\\ 
&+P_3\lambda_{(1,3)}^- - P_1\lambda_{(1,3)}^+\nonumber\\
0 =&-P_2\lambda_{(1,2)}^- + P_1\lambda_{(1,2)}^+ + P_3\lambda_{(2,3)}^- -P_2\lambda_{(2,3)}^+ \nonumber\\ 
&-P_2\lambda_{(0,2)}^- + P_0\lambda_{(0,2)}^+\nonumber\\
0 =&-P_3\lambda_{(2,3)}^- + P_2\lambda_{(2,3)}^+ - P_3\lambda_{(0,3)}^- +P_0\lambda_{(0,3)}^+ \nonumber\\ 
&-P_3\lambda_{(1,3)}^- + P_1\lambda_{(1,3)}^+.\nonumber\\
\end{align}
We choose $\lambda_{(n,m)}^- = 1-\lambda_{(n,m)}^+$, and solve the above set of equations to get
\begin{align}
\label{parameters_long}
\lambda_{(0,2)}^+ =& 
    \frac{2P_2-P_3}{P_0+P_2} 
    -\frac{P_1+P_2}{P_0+P_2}\lambda_{(1,2)}^+ 
    +\frac{P_2+P_3}{P_0+P_2}\lambda_{(2,3)}^+,\nonumber\\
\lambda_{(0,3)}^+ =& 
    \frac{2P_1-P_2+2P_3}{P_0+P_3}
    -\frac{P_0+P_1}{P_0+P_3}\lambda_{(0,1)}^+ \nonumber \\
    &+\frac{P_1+P_2}{P_0+P_3}\lambda_{(1,2)}^+ 
    -\frac{P_2+P_3}{P_0+P_3}\lambda_{(2,3)}^+,\nonumber\\
\lambda_{(1,3)}^+ =& -\frac{P_1+P_2-P_3}{P_1+P_3}
    +\frac{P_0+P_1}{P_1+P_3}\lambda_{(0,1)}^+
    -\frac{P_1+P_2}{P_1+P_3}\lambda_{(1,2)}^+,
\end{align} 
where the coefficients for close energy levels $\lambda^+_{(0,1)},\;\lambda^+_{(1,2)},\;\lambda^+_{(2,3)}$ are taken as free parameters. We now consider the same problem as if $\lambda^+_{(0,2)}=\lambda^+_{(0,3)}=\lambda^+_{(1,3)}=0$ (only nearest energy level transitions) and we get the simpler set of equations      
\begin{align}
0&=P_1\lambda_{(0,1)}^- - P_0\lambda_{(0,1)}^+ \nonumber\\
0&=-P_1\lambda_{(0,1)}^- + P_0\lambda_{(0,1)}^+ + P_2\lambda_{(1,2)}^- -P_1\lambda_{(1,2)}^+\nonumber\\
0&=-P_2\lambda_{(1,2)}^- + P_1\lambda_{(1,2)}^+ + P_3\lambda_{(2,3)}^- -P_2\lambda_{(2,3)}^+\nonumber\\
0&=-P_3\lambda_{(2,3)}^- + P_2\lambda_{(2,3)}^+  
\end{align}
which have the unique solution (by considering again $\lambda_{(n,m)}^+ = 1-\lambda_{(n,m)}^-$) 
\begin{align}
\label{parameters_short}
\lambda_{(0,1)}^+ &= \frac{P_1}{P_0+P_1},\nonumber\\
\lambda_{(1,2)}^+ &= \frac{P_2}{P_1+P_2},\nonumber\\
\lambda_{(2,3)}^+ &= \frac{P_3}{P_2+P_3}.
\end{align}
Using the results from Eq. (\ref{parameters_short}) in Eq. \eqref{parameters_long} we get the remaining parameters 
\begin{align}
\label{other_parameters_short}
\lambda_{(0,2)}^+ = \frac{P_2}{P_0+P_2},\nonumber\\
\lambda_{(0,3)}^+ = \frac{P_3}{P_0+P_3},\nonumber\\
\lambda_{(1,3)}^+ = \frac{P_3}{P_1+P_3}.
\end{align}
We thus have described a method to obtain the desired target density operator $\rho_T^{}$. It is simply necessary to rotate the operators from the two-site Hamiltonian diagonal basis to the computational basis to obtain the jump operators used in our calculations. We have given some details on how to implement this in Appendix \ref{sec:rotation_Gamma}.   

It is useful now to discuss about the target density operator $\rho_T^{}$ as various approaches can be followed. Here we use the two-site Hamiltonian 
\begin{eqnarray}
\label{local_hamiltonian}
H_{2S} &=& \frac{1}{2}\left[h_x(\sigma_1^x+\sigma_2^x)+h_z(\sigma_1^z+\sigma_2^z)\right]\nonumber\\
&&+ J(\sigma_1^x\sigma_2^x+\sigma_1^y\sigma_2^y) + \Delta\sigma_1^z\sigma_2^z.
\end{eqnarray}
which is the Hamiltonian for each internal bond of the chain of Eq.(\ref{XXZHamiltonian}). We then compute the thermal state $e^{-\beta H_{2S}}/\tr(e^{-\beta H_{2S}})$ and diagonalize it.   
%
\begin{figure}[t]
\begin{center}
\includegraphics[width=\linewidth]{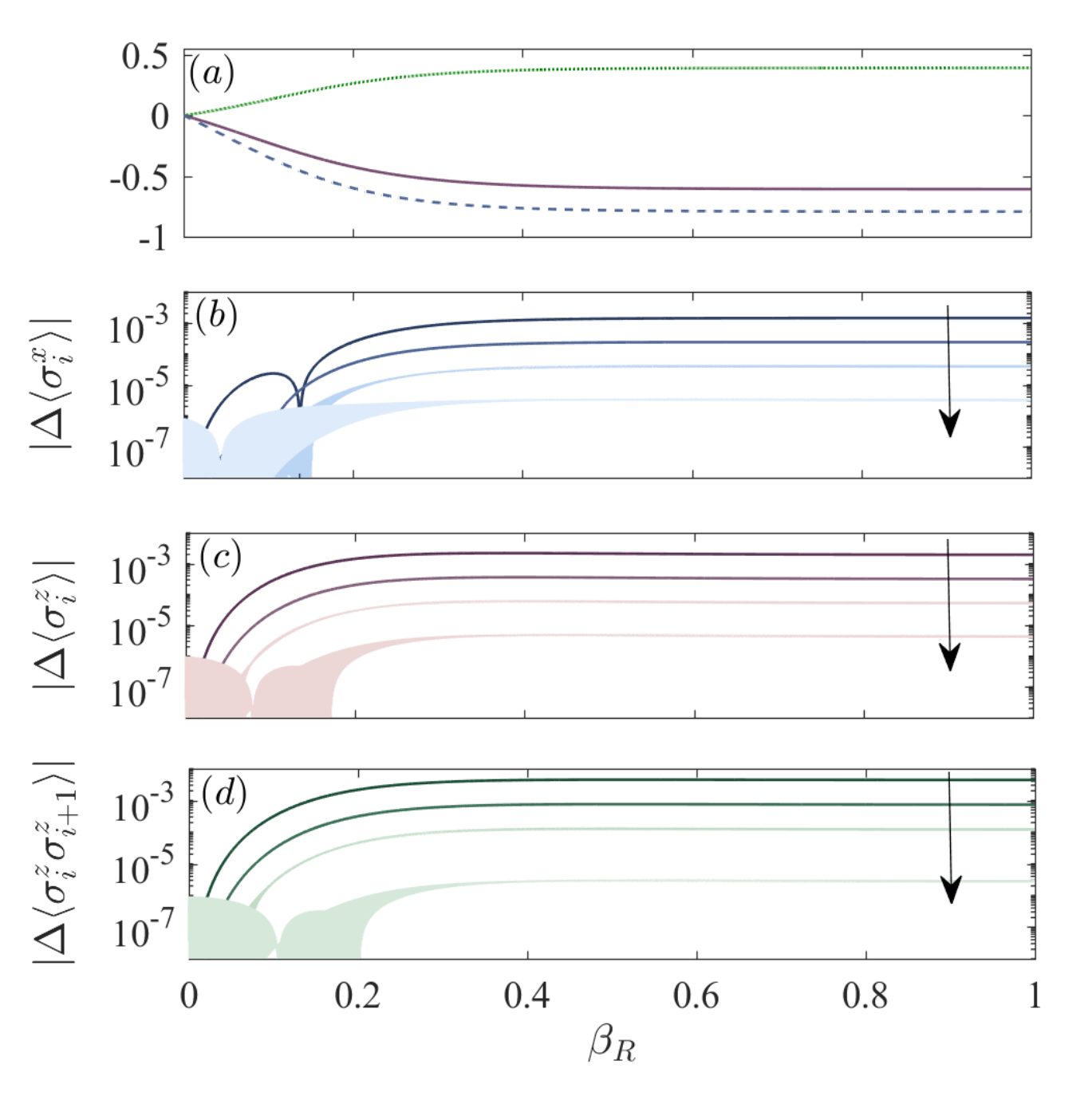}
\caption{\label{fig:fig1} $(a)$ $\langle \sigma^x_i \rangle$ (blue dashed line), $\langle \sigma^z_i \rangle$ (red solid line) and $\langle \sigma^z_i\sigma^z_{i+1} \rangle$ (green dotted line), versus inverse temperature of the thermal state $\beta$ for a chain length $L=11$. 
$(b$-$d)$ Absolute value of the difference between observables computed for, in the order indicated by the direction of the arrow, $L=7$ and $L=5$, $L=9$ and $L=7$, $L=11$ and $L=9$, $L=13$ and $L=11$. In particular, in panel $(b)$ we show $|\langle \sigma^x_i \rangle|$, in $(c)$ $|\langle \sigma^z_i \rangle|$, and in $(d)$ $|\langle \sigma^z_i \sigma^z_{i+1} \rangle|$.        
In all panels $i=(L+1)/2$, $h_x=3.375$, $h_z=2$, $J=-0.5$ and $\Delta=-1$.        
}
\end{center}
\end{figure}
%

%
%
\section{Results} \label{sec:results}     
In this section we study the steady state that is reached when applying the dissipator discussed in Sec.~\ref{ssec:our}. If the steady state is an exact thermal state, then the expectation value of any observable is determined once the temperature and the Hamiltonian are known. Consequently, by evaluating different observables, it is possible to check whether their expectation values are consistent with the steady state being thermal. In the following we consider different observables and we compute their expectation values in the steady state. Then, for each of them we check what would be the temperature of a thermal state which has the same expectation value for that observable. To do so, however, it is first necessary to have a reference temperature for the expectation values that the various observables can take. We discuss this in Sec.~\ref{sec:reference}. We will then consider both static and dynamical observables, respectively in Sec.~\ref{sec:1time} and Sec.~\ref{sec:2time}. Hereafter, for simulations with $L \ge 13$ we use a matrix product states algorithm for open systems \cite{Schollwock2011, VerstraeteCirac2004, bonddimension}.

\subsection{Reference temperature}\label{sec:reference}        

From the Hamiltonian in Eq. (\ref{XXZHamiltonian}), we compute thermal states for the reference inverse temperature $\beta_R\in [0,1]$ for chains of lengths from $L=5$ to $13$. The typical behavior of observables as a function of $\beta_R$ is depicted in Fig. \ref{fig:fig1}$(a)$. In particular it shows $\langle\sigma^z_l\rangle$, $\langle\sigma^x_l\rangle$ and $\langle\sigma^{z}_l\sigma^{z}_{l+1}\rangle$ with $l=6$, the middle site of a chain with length $L=11$.    
At low $\beta_R$ (high temperatures) the observable changes sizeably with $\beta_R$, while at larger $\beta_R$ (lower temperatures), the observable is almost constant and it would be very hard to clearly identify the temperature corresponding to a given value of the observable. 
In Figs. \ref{fig:fig1}$(b$-$d)$ we consider five different chain lengths $L(k)=2k+3$ with $k\in[1,5]$, and in each panel we show the absolute value of the difference between the expectation value of an observable in the middle of the chain for two different lengths $L(k+1)$ and $L(k)$, plotted versus the reference inverse temperature $\beta_R$. For example, in Fig. \ref{fig:fig1}$(b)$ we show $|\Delta \langle \sigma^x_i\rangle | = |\langle\sigma^x_{k+3}\rangle_{k+1,\beta_R} - \langle\sigma^x_{k+2,\beta_R}\rangle_{k,\beta_R}|$ where with $\langle\cdot\rangle_{k,\beta_R}$ we mean the expectation value for a chain of length $L(k)$ and at an inverse temperature $\beta_R$. 
In Fig. \ref{fig:fig1}$(c)$ we show $|\Delta \langle \sigma^z_i\rangle | = |\langle\sigma^z_{k+3}\rangle_{k+1,\beta_R} - \langle{\sigma^z_{k+2}}\rangle_{k,\beta_R}|$ and in Fig. \ref{fig:fig1}$(d)$ we show $|\Delta \langle \sigma^z_i\sigma^z_{i+1}\rangle | = |\langle\sigma^z_{k+3}\sigma^z_{k+4}\rangle_{k+1,\beta_R} - \langle\sigma^z_{k+2}\sigma^z_{k+3}\rangle_{k,\beta_R}|$.  
Figs. \ref{fig:fig1}$(b$-$d)$ show that for large enough temperatures and for $L=11$ the value of the observable in the middle of this non-integrable chain has converged. It is thus possible to use a chain of size $L=11$ to associate a temperature to a certain value of an observable. 

In the following we use Eq.(\ref{Lindbladian}) to derive a steady state. Then, given and observable at a site $i$, we associate an effective inverse temperature which is given by the reference inverse temperature $\beta_R$ of a thermal state for which the expectation value of the same observable is identical. For example, when studying a chain of length $L=17$, we compute $\langle \sigma^x_9 \rangle$ and check for which inverse temperature $\beta_R$ a thermal chain would have the same value for this observable. We then refer to this inverse temperature as $\beta_x$. Analogously we compute $\beta_z$ and $\beta_{zz}$. 
We shall note here that in \cite{Mendoza-ArenasJaksch2015} the authors suggest to compute the effective temperature by the minimum distance between a local reduced density matrix and the one from a reference thermal state.                             

\subsection{Static observables}\label{sec:1time}     
%
%

%
\begin{figure}[t]
\begin{center}
\includegraphics[width=\linewidth]{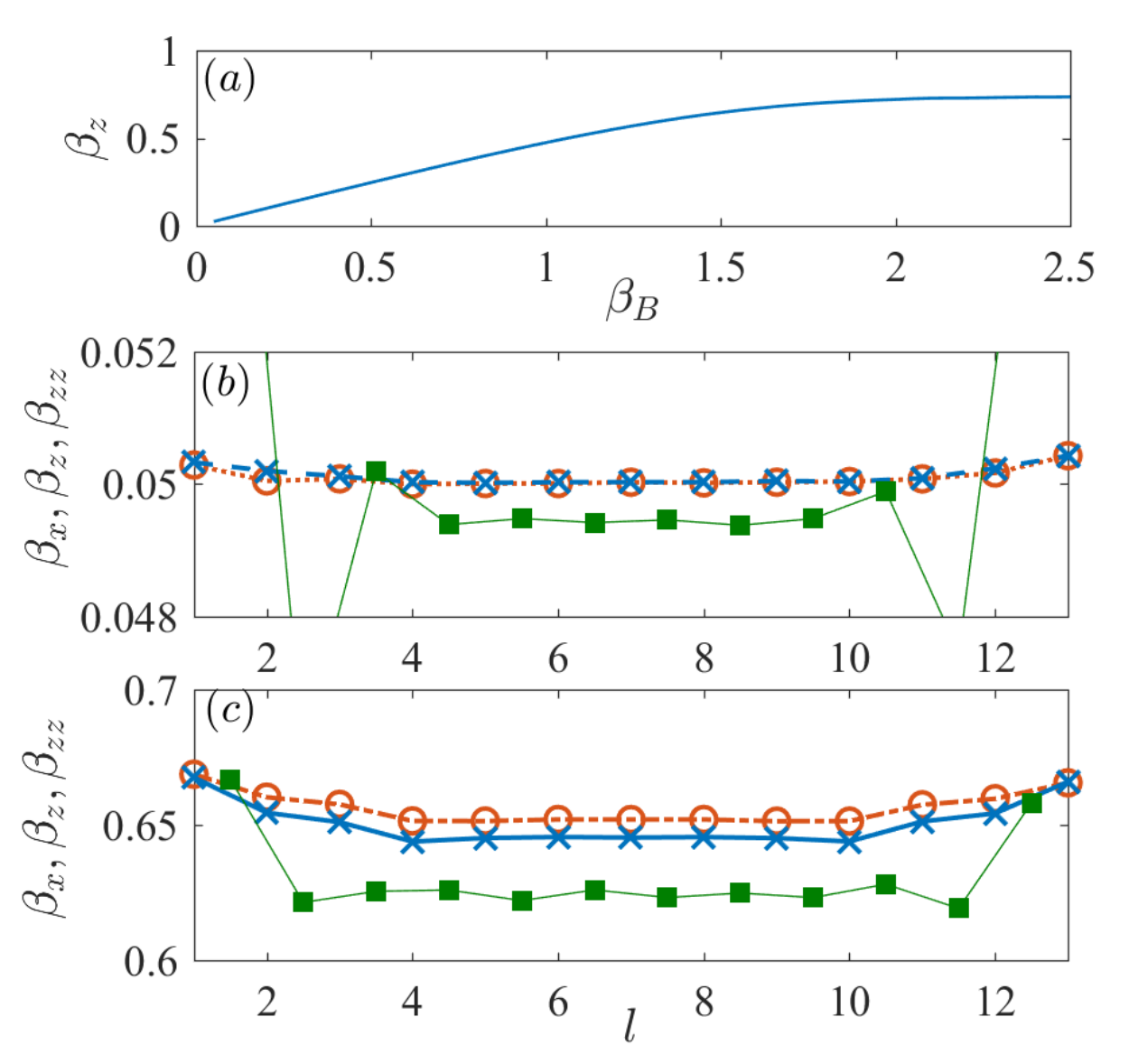}
\caption{\label{fig:fig2} $(a)$ Effective inverse temperature $\beta_z$ associated with the expectation value $\langle \sigma^z_7\rangle$ (middle of the chain) versus temperature of the dissipative bath $\beta_B$. $(b,c)$ Effective inverse temperature $\beta_x,\;\beta_z,\;\beta_{zz}$ respectively for the dashed blue line with $\times$, dot-dashed red line with $\circ$ and solid green line with $\square$. In panel $(b)$ $\beta_B=0.1$ while in panel $(c)$ $\beta_B=1.5$. In all panels $L=13$, $h_x=3.375$, $h_z=2$, $J=-0.5$, $\Delta=-1$ and $\gamma=0.35$.        
}
\end{center}
\end{figure}
%

We first consider the effective inverse temperatures $\beta_z$, computed from the middle site of the chain, versus the target inverse temperature $\beta_B$ that the dissipator $\mathcal{D}$ is trying to impose at the edge. In Fig. \ref{fig:fig2}$(a)$ we show $\beta_z$ versus $\beta_B$. For $\beta_B\lesssim 1.5$ we find that $\beta_z$ grows linearly, whereas for larger values of $\beta_B$ it reaches a plateau. We have observed a similar behavior for all observables examined and for different types of chain Hamiltonians (here we show the XXZ chain with $h_x=3.375$, $h_z=2$, $J=-0.5$, $\Delta=-1$ and $L=17$).    
We then analyze the effective temperature reached in the system at different sites for low bath inverse temperature, $\beta_B=0.1$ in panel $(b)$, and larger bath inverse temperature, $\beta_B=1.5$ in panel $(c)$. As discussed previously, the effective temperature associated at each site to an observable is given by the temperature of a thermal state for which that observable has the same expectation value, e.g., $\beta_x$ is associated to $\langle\sigma^x_l\rangle$, $\beta_z$  to $\langle\sigma^z_l\rangle$ and  $\beta_z$ to $\langle\sigma^z_l\sigma^z_{l+1}\rangle$. The difficulty here is that the size of the systems taken as reference is limited (here we consider $L=11$ for the reference thermal state). Nevertheless, the value of the observables in a non-integrable chain in a thermal state converges to a certain value within a few sites from the edges of the chain. Hence we can use the values at the edge of the reference thermal chains to obtain the effective temperature at the edge of the chain from the master equation, and for the other sites we use the bulk values. Note that in order to have accurate reference values for the inverse temperature, we use $10^5$ values of $\beta_R$.         

In Figs. \ref{fig:fig2}$(b,c)$ we plot the effective temperature for $\beta_z$ (red dot-dashed line with $\circ$), $\beta_x$ (blue dashed line with $\times$) and $\beta_{zz}$ (green solid line with $\square$). Note that here the dissipator $\mathcal{D}$ is acting only on sites $l=1,2$. At lower inverse temperature, panel $(b)$, the inverse temperatures $\beta_z$ and $\beta_x$ are practically indistinguishable for $l\geq 3$, and the temperature is fairly constant throughout the bulk of the chain. For an effective inverse temperature related to the correlation $\beta_{zz}$, we find that in the bulk there is an error of only a few percents, while at the edge the difference is much greater. At larger inverse bath temperature $\beta_B=1.5$, see Fig.~\ref{fig:fig2}$(c)$, the effective local inverse temperatures are much more clearly different from each other, highlighting the difficulty in reaching good thermal states at larger inverse temperatures.     
%
\begin{figure}[t]
\begin{center}
\includegraphics[width=\linewidth]{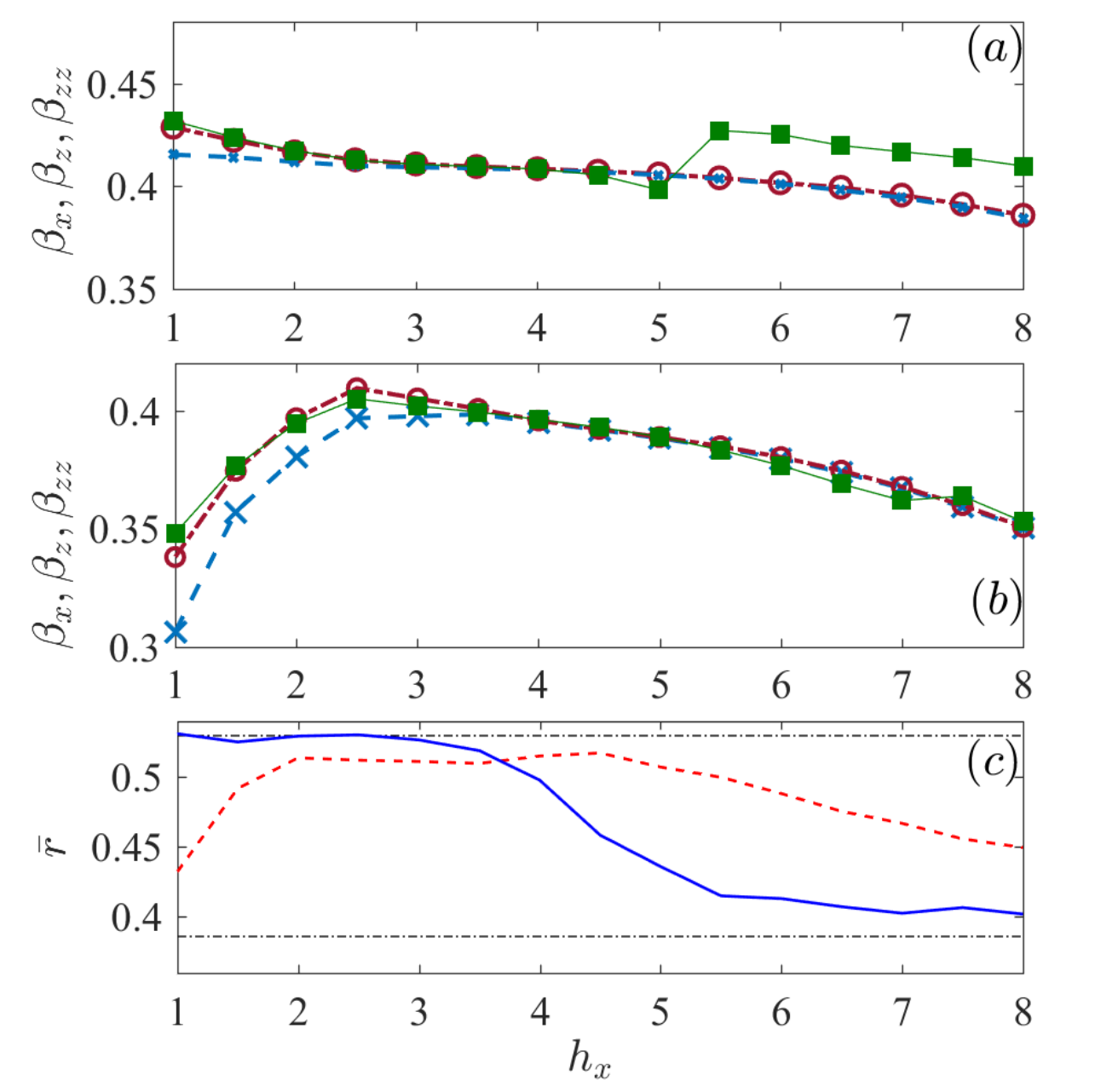}
\caption{$(a,b)$ Effective temperature for different observables as a function of the transverse magnetic field $h_x$ which breaks the integrability of the chain. In $(a)$ we consider an XXZ chain with $J=-0.5$, while in $(b)$ we consider an Ising chain, i.e., $J=0$. In all panels $L=17$, $h_z=2$ and $\Delta=-1$. In $(a,b)$ we have used $\gamma=0.35$. $(c)$ Mean value $\bar{r}$ versus transverse field $h_x$ for $J=-0.5$ and $L=15$ (blue solid line), and for $J=0$ and $L=16$ (red dashed line). The dot-dashed thin black lines represent the values $\bar{r}=0.386$ and $0.5295$. \label{fig:fig3}
}
\end{center}
\end{figure}
%
As mentioned briefly earlier, it is expected that the chain reaches a steady state which resembles more a thermal state when it is non-integrable. Fig. \ref{fig:fig3} confirms this understanding. In both panel $(a)$ and panel $(b)$ we plot the inverse temperature for different observables $\beta_M$ against the Hamiltonian parameter $h_x$. In both panels we use $h_z=2$, $\Delta=-1$, $L=17$ and $\beta_B=0.85$, while in panel $(a)$ $J=-0.5$ and in panel $(b)$ $J=0$. We observe that in some regions the inverse temperatures of the different observables are close to one another, or even indistinguishable in these plots, while in other regions the inverse temperatures are very different. With Fig. \ref{fig:fig3}$(c)$ we can compare the behavior of the effective inverse temperatures with the variation of the mean $\bar{r}$ as a function of $h_x$. In particular, the blue solid line shows $\bar{r}$ for $L=15$, $h_z=2$, $\Delta=-1$ and $J=-0.5$, while $L=16$ and $J=0$ for the red dashed line. The black dot-dashed horizontal lines show the values for $\bar{r}=0.386$ which would be obtained from a random Hamiltonian with the energy level difference following a Poisson distribution, and $\bar{r}=0.5295$ which is computed from large Gaussian orthogonal ensemble matrices. Although we do not observe a clear quantitative match between $\bar{r}$ and the proximity of the effective inverse temperatures, comparing Fig. \ref{fig:fig3}$(c)$ with Fig. \ref{fig:fig3}$(a,b)$ we observe that when $\bar{r}$ is largest, the inverse temperatures match better between each other, compared to when $\bar{r}$ is closer to $0.386$. In Fig. \ref{fig:fig3}$(a)$ we see that for $h_x\lesssim 4$ the inverse temperature are closer than for $h_x\gtrsim 4$, which is consistent with the change in $\bar{r}$ shown in Fig. \ref{fig:fig3}$(c)$ by the continuous blue line. In Fig. \ref{fig:fig3}$(b)$ we zoom-in on the $y-$axis to see more clearly that the region of parameters of the transverse field $h_x$ in which the inverse temperatures match the most is approximately for $h_x\in (3,5)$, which is the region in which $\bar{r}$ is largest (red dashed line in Fig. \ref{fig:fig3}$(c)$).    

%
\begin{figure}[t]
\begin{center}
\includegraphics[width=\linewidth]{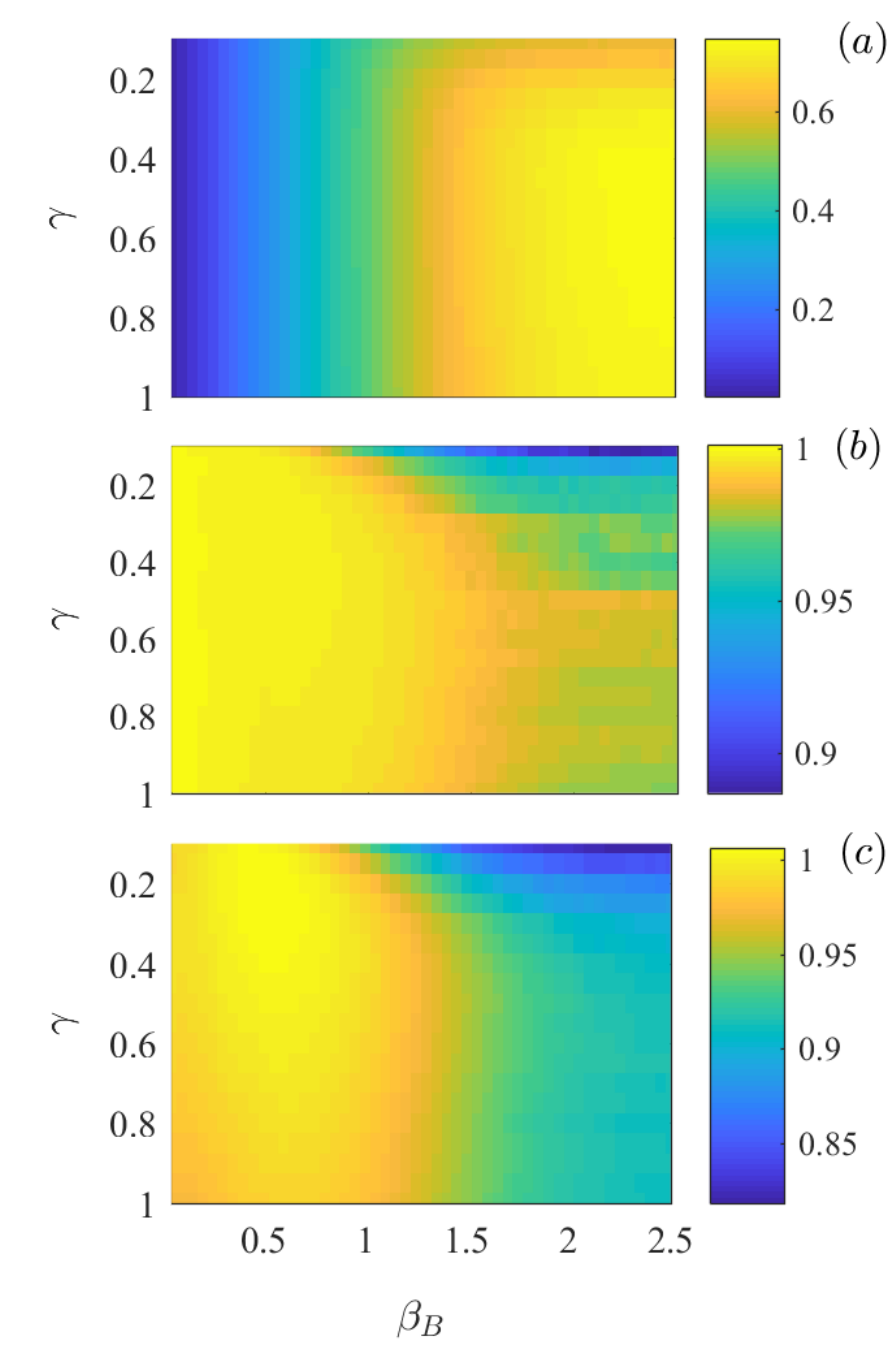}
\caption{$(a)$ $\langle \sigma^z_i\rangle$ as a function of the strength of the coupling to the bath $\gamma$ and of the bath inverse temperature $\beta_B$ for $i=(L+1)/2$. The observable does not depend significantly on $\gamma$, especially for low inverse temperature $\beta_B$. (b-c) Relative temperature of the observable $\beta_x/\beta_z$ $(b)$ and $\beta_{zz}/\beta_z$. For inverse temperatures $\beta_B\lesssim 1$ the ratio between the temperatures is close to $1$. In all panels $L=17$, $h_x=3.375$, $h_z=2$, $J=-0.5$ and $\Delta=-1$. \label{fig:scan}}
\end{center}
\end{figure}
%
In order to find the best scenarios in which the steady state is closer to a thermal state, it is also important to check the role of the dissipation rate $\gamma$. In Fig. \ref{fig:scan} we show intensity plot of $\beta_z$, panel $(a)$, $\beta_x/\beta_z$, panel $(b)$, and $\beta_{zz}/\beta_z$, panel $(c)$, versus inverse bath temperature $\beta_B$, and dissipation ratio $\gamma$. Fig. \ref{fig:scan}$(a)$ shows that when $\gamma$ is large enough (here e.g., for $\gamma \gtrsim 0.3$), the resulting inverse temperature $\beta_z$ shows little dependence on $\gamma$ itself. The behavior of the inverse temperature from other observables, $\beta_x$ and $\beta_{zz}$, with $\gamma$ has a similar behavior as $\beta_z$, as can be inferred from Figs. \ref{fig:scan}$(b,c)$.  
Going back to Fig. \ref{fig:scan}$(a)$, the dependence on $\beta_B$, for a given $\gamma$, seems to reach a plateau for $\beta_B\gtrsim 1.5$. In Fig. \ref{fig:scan}$(b,c)$ we note that the ratio between the inverse temperatures is closer to $1$ for larger temperatures, however the dependence of $\beta_{zz}/\beta_z$ with $\beta_B$ is, in general, non-monotonous.         

%
\begin{figure}[t]
\begin{center}
\includegraphics[width=\linewidth]{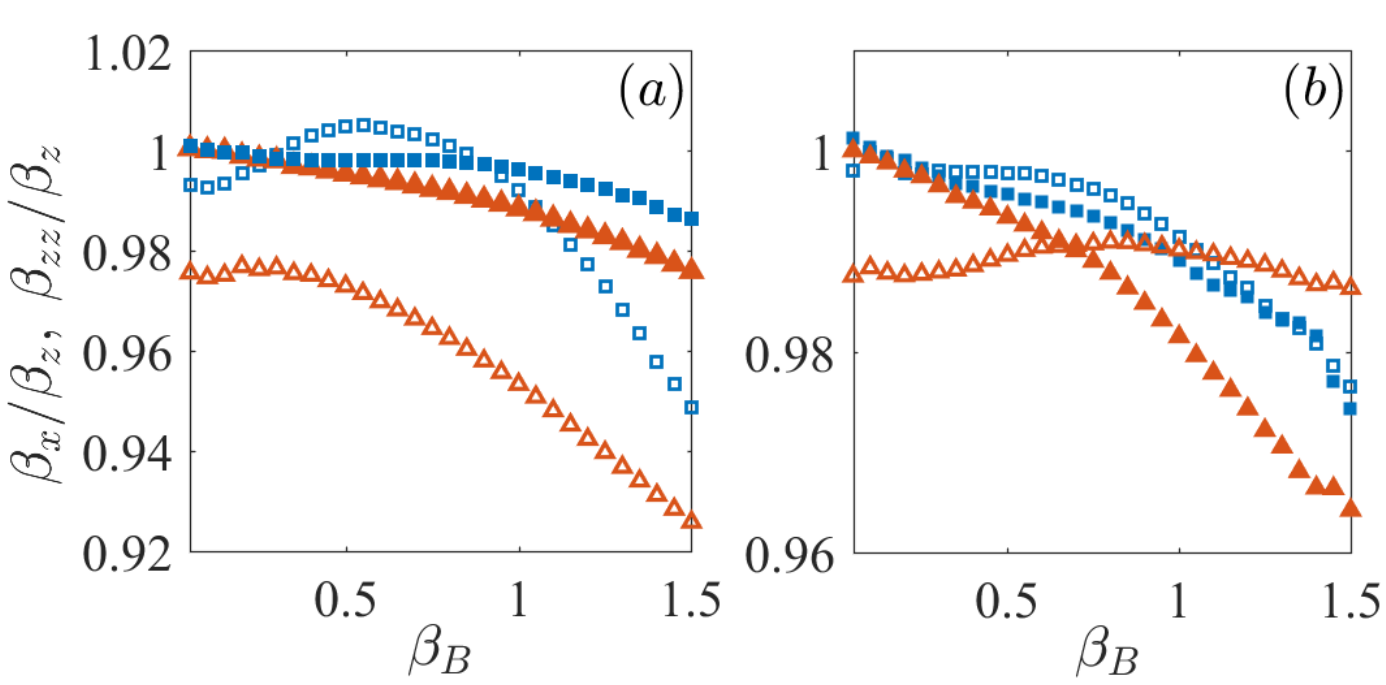}
\caption{Comparison between the obtained effective temperatures for the detailed balance (blue squares) and the fast relaxation (red triangles) approaches versus $\beta_B$. In $(a)$ we consider an $XXZ$ chain, with $J=-0.5$, while in (b) we have an Ising chain with $J=0$. The full symbols represent $\beta_x/\beta_z$, while the empty marks represent $\beta_{zz}/\beta_z$. In both scenarios, the detailed balance approach performs better than the fast relaxation one. In all panels $L=17$, $h_x=3.375$, $h_z=2$, $\Delta=-1$ and $\gamma=0.35$. \label{fig:comparison}  
}
\end{center}
\end{figure}
%
To evaluate the performance of the detailed balance approach, we now compare it to the fast relaxation one. In Fig. \ref{fig:comparison}, we plot the ratio of the inverse temperatures for different observables for both methods and for different non-integrable chains. More specifically we use blue squares for the detailed balance approach and red triangles for the results from fast relaxation. In Fig. \ref{fig:comparison} we depict $\beta_x/\beta_z$ with filled symbols and $\beta_{zz}/\beta_z$ with empty ones. In Fig. \ref{fig:comparison}$(a,b)$, we consider chains with $h_x=3.375$, $h_z=2$, $\Delta=-1$, $L=17$ and $\gamma = 0.35$ while in Fig. \ref{fig:comparison}$(a)$ $J=-0.5$ and in Fig. \ref{fig:comparison}$(b)$ $J=0$. We have chosen to represent these two cases as they are fairly representative of what we have observed. In particular, the case in Fig. \ref{fig:comparison}$(b)$ is one of the best we have found for the fast relaxation approach, while in Fig. \ref{fig:comparison} we have chosen a set of parameters which gives the best performance for the detailed balance approach. In both cases, for the detailed balance approach both inverse temperature ratios are within few percents of $1$ for the inverse temperature of the bath $\beta_B\lesssim 1$. The ratios of inverse temperatures are in general further away from $1$ for the fast relaxation approach, and in particular the ratio of the inverse temperatures $\beta_{zz}/\beta_z$ is further away from $\beta_x/\beta_z$ compared to the detailed balance approach. It should be pointed out that the regime of parameters $\gamma$ and $\beta_B$ for which the relative error between the inverse tempratures is within $1\%$ is much broader for the detailed balance approach compared to the fast relaxation one. Moreover, the detailed balance approach can get simultaneously well bellow the $1\%$ threshold for both ratios.
We should add that while reaching the steady state is faster, for all cases we checked, with the fast relaxation method, the time required by detailed balance approach to reach the steady state approaches that of the fast relaxation method when we increase the size of the system. Hence the improved accuracy of the detailed balance approach does not come at a particularly high cost in simulation time. It should also be mentioned that while we consider here chains of intermediate size, the ratio between effective temperatures from different observables does not improve significantly when increasing the system size.  

\subsection{Dynamical correlations}\label{sec:2time}    
%
%
%
\begin{figure}[t]
\begin{center}
\includegraphics[width=\linewidth]{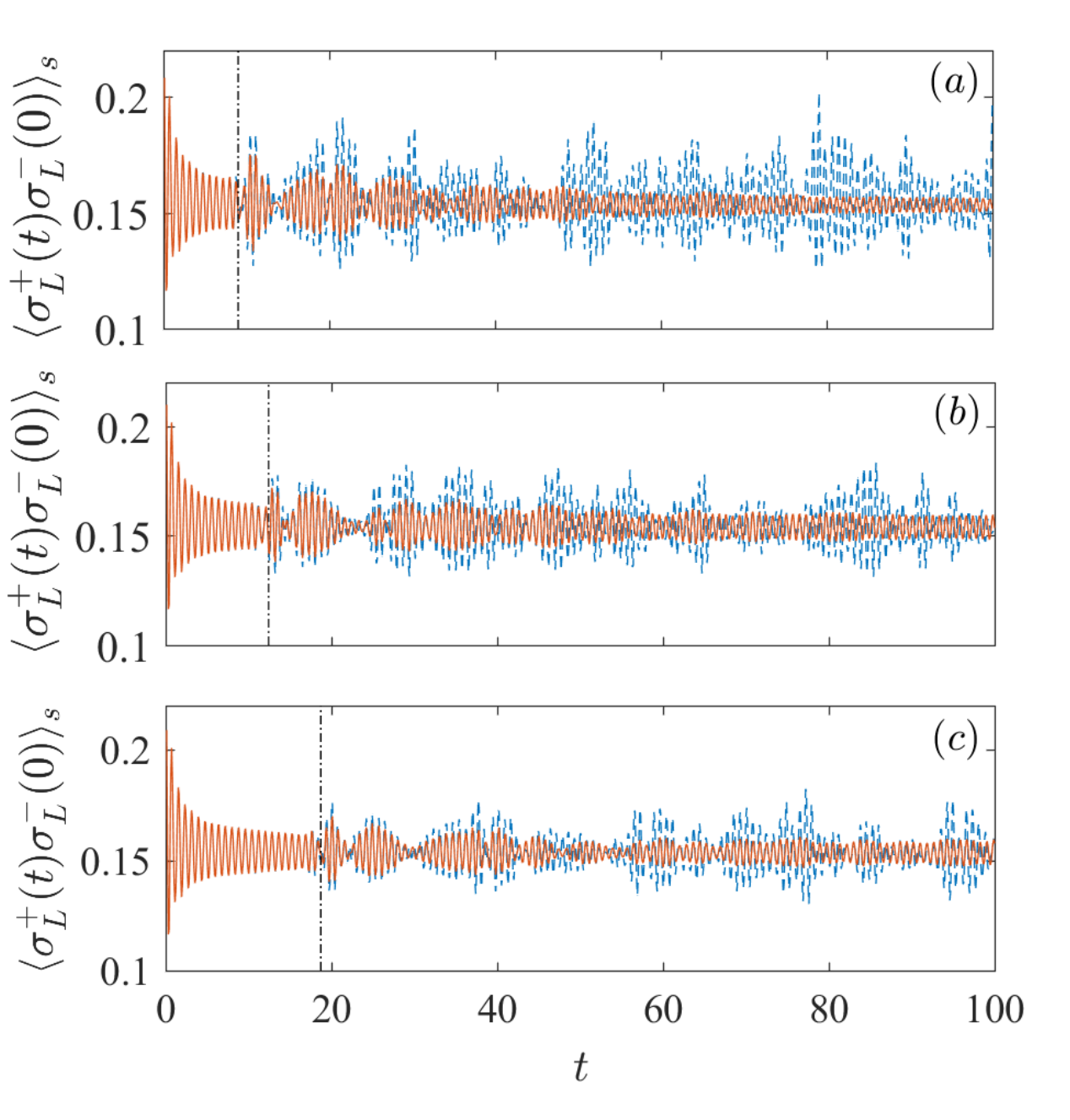}     
\caption{\label{fig:fig6} Two-time correlations $\langle\sigma_1^+(t)\sigma_1^-(0)\rangle$ versus time $t$ in the side not in contact with the bath, for the (almost thermal) steady state when the bath is still acting on the chain ($\gamma=0.35$ red solid lines), or when the bath is decoupled from the chain after it has reached the steady state ($\gamma=0$ blue dashed line). In panels (a-c) we consider different length chains: in $(a)$ $L=13$, in $(b)$ $L=17$ and in $(c)$ $L=25$. The dashed vertical black line represent the time at which boundary effects first come into play. In all panels , $h_x=3.375$, $h_z=2$, $J=-0.5$, $\beta_B = 0.85$ and $\Delta=-1$.              
}
\end{center}
\end{figure}
%
%
\begin{figure}[t]
\begin{center}
\includegraphics[width=\linewidth]{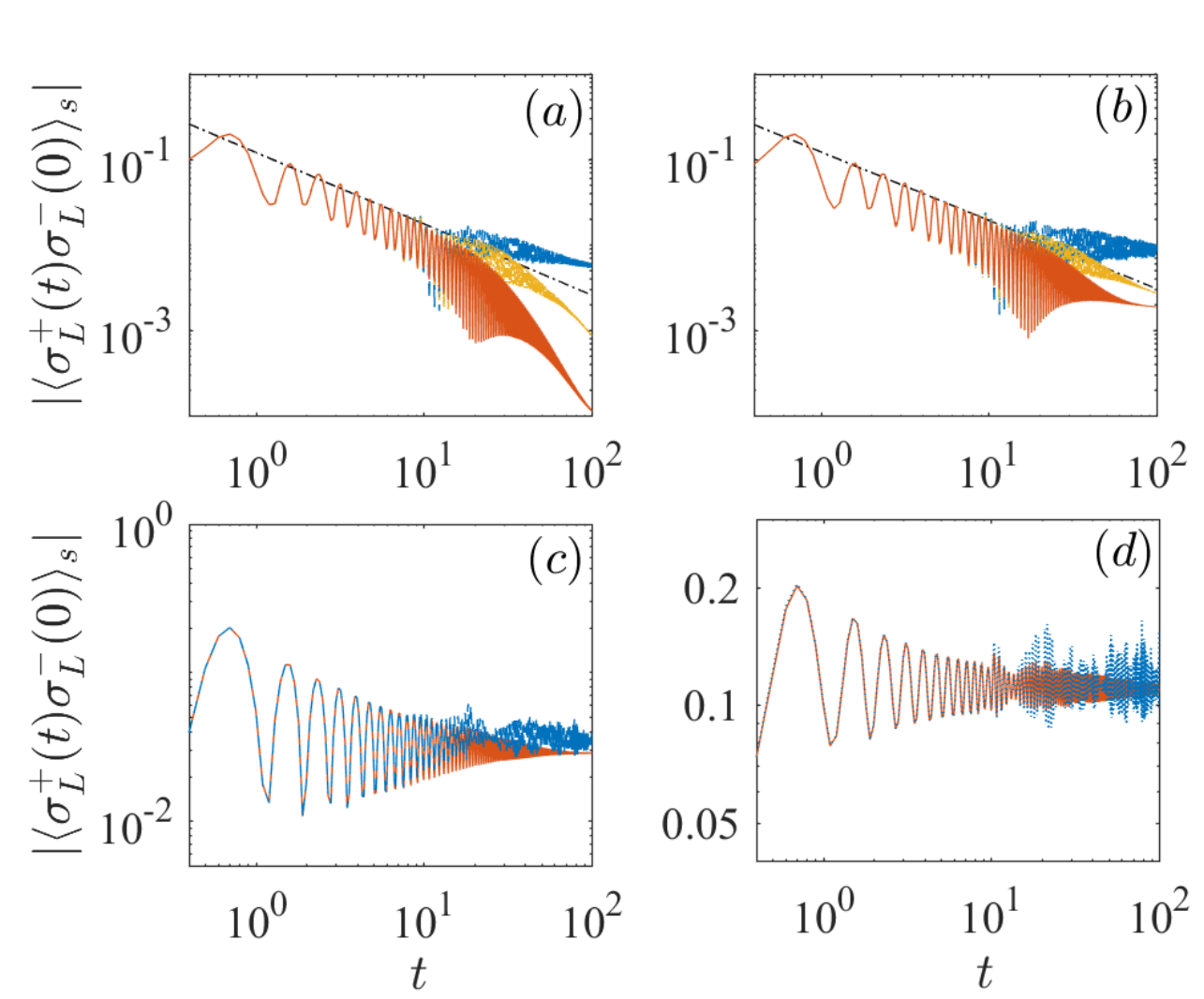}
\caption{\label{fig:fig7} Absolute values of the two-time correlations $\langle\sigma_1^+(t)\sigma_1^-(0)\rangle$ versus time $t$, in the side not in contact with the bath, for the (almost thermal) steady state. We consider different effective observable temperatures $\beta_B=0.01$ $(a)$, $\beta_B=0.05$ $(b)$,  $\beta_B=0.2$ $(c)$ and  $\beta_B=0.5$ $(d)$. We show the two-time correlations when the bath is still acting on the chain with $\gamma=0.35$ as in the preparation of the state (red solid lines), or when the bath is decoupled from the chain after it has reached the steady state (blue dashed line). The dot-dashed straight lines in $(a)$ and $(a)$ are power-law fits to the maxima of the two-time correlations. In panels $(a)$ and $(b)$ the yellow line shows the two time correlation for $\gamma=0.1$. In all panels $h_x=3.375$, $h_z=2$, $J=-0.5$ and $\Delta=-1$. 
}
\end{center}
\end{figure}
%
While it is possible to produce steady states with targeted static properties, their dynamical properties may differ significantly from those of the target state \cite{SciollaKollath2015}. For this reason we investigate the two-time correlation 
\begin{align}
\langle \sigma^+_i(t) \sigma^-_i(0) \rangle_s = \tr\left[ \sigma^+_i e^{\mathcal{L}t}\left(\sigma^-_i \rho_s\right)\right] \label{eq:twotimecorr}
\end{align}
where $\langle \dots\rangle_s$ stresses that the two-time correlation is computed from the steady state $\rho_s$. In order to evaluate such a quantity, we first compute the steady state of Eq. (\ref{Lindbladian}), we then apply $\sigma^-_i$ to it, followed by evolving the system with Eq. (\ref{Lindbladian}). Last we apply $\sigma^+_i$ and take the trace.  

We now consider the spin chain Hamiltonian of Eq. (\ref{XXZHamiltonian}) and we couple the two-site bath with the parameters from the detailed balance approach only on one edge of the chain (sites $1$ and $2$). We then study the two-time correlation $\langle \sigma^+_L(t) \sigma^-_L(0) \rangle_s$ on the last site for different system lengths $L$, bath inverse temperatures $\beta_B$ and stronger coupling strengths $\gamma$. In Fig. \ref{fig:fig6} we consider different system size, in particular $L=13$ for panel $(a)$, $L=17$ for panel $(b)$ and $L=25$ for panel $(c)$. In each panel we show two curves: the red solid line represents the two-time correlations while keeping the system coupled to the Lindblad bath (i.e., $\gamma\ne 0$), while for the blue dashed line we take $\gamma=0$, i.e., the chain which was thermalized by the bath is now no longer coupled to it \cite{fn1}. We notice that at a certain time, indicated in the figures by a dot-dashed vertical line, the effects of the boundaries, in the form of beating-like oscillations, are felt in the correlation. In fact, the time at which the two-time correlation starts to have beatings increases with the system size. In absence of coupling to the bath the finite size effects are much larger, especially for shorter chains. 

In Fig. \ref{fig:fig7} we study the evolution of the absolute values of the two-time correlations for different values of the inverse temperatures of the bath $\beta_B$. Similarly to Fig. \ref{fig:fig6} the red solid lines correspond to an evolution in which the chain and the bath are always connected, while the blue dashed lines (dotted line in panel $(d)$) are for evolutions with $\gamma=0$. The inverse temperature increases in the panel from $\beta_B=0.01$ in panel $(a)$, $\beta_B=0.05$ in panel $(b)$, $\beta_B=0.2$ in panel $(c)$ and $\beta_B=0.5$ in panel $(d)$. At larger $\beta_B$, the correlations are expected to decay more slowly, however, in this finite-size chain, the correlations seem to decay only for a finite time, and then stabilize to a non-zero value for long times (as expected the oscillations around the mean are smaller when $\gamma\ne 0$). 
At lower inverse temperatures, the correlations are expected to decay faster. We see that the correlations at the beginning follow a decay well approximated by a power law. At longer time the finite-size effects come into play. For $\gamma=0$ the correlations decay more slowly, while if the chain is still coupled to the bath, i.e., $\gamma= 0.35$ (red solid lines), the correlations can decay even faster than the initial power-law regime. Using an intermediate strength of coupling to the bath, see yellows lines for $\gamma=0.1$ in Fig. \ref{fig:fig7}$(a,b)$, results in a decay that resembles the power-law decay for longer times.

\section{Conclusions}\label{sec:conclusions}      
We have presented an alternative method to produce thermal states with a two-site GKSL master equation for non-integrable systems. We have characterized the performance of the method by analysing the expectation values of single and two-site observables in the steady state, and we found that different observables can be consistent with those from a thermal state as long as we accept a relatively small error on its temperature. 

We have also studied dynamical correlations, which give a clear indication, depending on the temperature, of how long a chain should be for its dynamical properties to look thermal enough. As expected, for larger temperature the chain can be shorter, while for smaller temperatures a much longer chain is required. The details depend on the particular parameters of the Hamiltonian.

Since it is possible to define many GKSL master equations to obtain the same target density operator $\rho_T^{}$, further research could be focused on optimization of the parameters of the Hamiltonian, and/or of the dissipator to reach better thermal states. Here we have used, as target density operator, the thermal state of the Hamiltonian $H_{2S}$ in Eq. (\ref{local_hamiltonian}), however, for the fast relaxation method other target $\rho_T^{}$ have been used as, for example, the thermal state for the first two sites of the Hamiltonian in Eq. (\ref{XXZHamiltonian}) \citep{Mendoza-ArenasJaksch2015}, or the reduced density operator of the middle two sites of a long enough thermal chain with Hamiltonian (\ref{XXZHamiltonian}) \cite{Znidaric2011}. When different target states are chosen, the relation between the effective inverse temperature obtained in the system (e.g., $\beta_x$) and the one of the bath $\beta_B$ will be quantitatively different from the one represented in Fig. \ref{fig:fig2}$(a)$.    
We have run various tests using also the $\rho_T^{}$ described in \cite{Mendoza-ArenasJaksch2015,Znidaric2011}, and we found that using Eq. \ref{local_hamiltonian} was consistently resulting in ratios of effective inverse temperatures closest to $1$. However a thorough investigation is required to shed light on which would be the best target state $\rho^{}_T$.

{\bf Acknowledgments}: We acknowledge fruitful discussions with M. \v{Z}nidari\v{c}. D.P. and M.P. acknowledge support from the Singapore Ministry of Education, Singapore Academic Research Fund Tier-II (project MOE2016-T2-1-065).

\begin{appendix} 
  \section{Representation of the jump operators in the computational basis} \label{sec:rotation_Gamma}  
We describe here how we write the jump operators in a way that it is conveniently implemented in our code. The operators which are implemented are the raising operator $\sigma^+=|\ua\rangle\langle\da|$, the lowering operator $\sigma^-=|\da\rangle\langle\ua|$, the projector on the $|\ua\rangle$ state $\sigma^u=|\ua\rangle\langle\ua|$, and the projector on the $|\da\rangle$ state $\sigma^d=|\da\rangle\langle\da|$. We now write the density operator as 
\begin{align}
\rho_T &= \sum_{s_1,s_2,t_1,t_2} \rho^{s_1,s_2}_{t_1,t_2}|s_1, s_2\rangle\langle t_1, t_2| \nonumber \\
&= V \left( \sum_{j=0}^3 P_j |j\rangle\langle j| \right)V^\dagger    
\end{align}
where 
\begin{align}
V = \sum_{s_1,s_2,j} \mu^{s_1,s_2}_{j} |s_1, s_2\rangle\langle j|     
\end{align}
and where $s_1$, $s_2$, $t_1$, $t_2$ can take the values $\ua$ or $\da$. The $\mu^{s_1,s_2}_j$ are given by diagonalizing $\rho^{s_1,s_2}_{t_1,t_2}$ (e.g., after having combined the two indices $s_1$ and $s_2$ in a single one). 
The jump operators in the computational basis are then given by   
\begin{align}
V \;\Gamma_{(n,m)} V^\dagger = \sum_{s_1,s_2,t_1,t_2} \mu^{s_1,s_2}_{p} (\mu^{t_1,t_2}_{q})^* |s_1, s_2\rangle\langle t_1, t_2|.           
\end{align}
In order to get the jump operators for the code, we need to translate $|s_1, s_2\rangle_l\langle t_1, t_2|$ to lowering/raising/projection operators, which can be done, for example, from 
\begin{align}
\left( \begin{array}{c}
|\da, \da\rangle \\ 
|\da, \ua\rangle \\ 
|\ua, \da\rangle \\ 
|\ua, \ua\rangle  
\end{array} \right) 
= \left(\begin{array}{cccc}
\sigma^d_1\sigma^d_2 & \sigma^d_1\sigma^-_2 & \sigma^-_1\sigma^d_2 & \sigma^-_1\sigma^-_2 \\ 
\sigma^d_1\sigma^+_2 & \sigma^d_1\sigma^u_2 & \sigma^-_1\sigma^+_2 & \sigma^-_1\sigma^u_2 \\ 
\sigma^+_1\sigma^d_2 & \sigma^+_1\sigma^-_2 & \sigma^u_1\sigma^d_2 & \sigma^u_1\sigma^-_2 \\ 
\sigma^+_1\sigma^+_2 & \sigma^+_1\sigma^u_2 & \sigma^u_1\sigma^+_2 & \sigma^u_1\sigma^u_2 
\end{array}\right) 
\left(\begin{array}{c}
|\da, \da\rangle \\ 
|\da, \ua\rangle \\ 
|\ua, \da\rangle \\ 
|\ua, \ua\rangle  
\end{array} \right),        
\end{align} 
where $\sigma^a_l$ (with $a=u,d,+,-$) acts on site $l$. 
Since the master equation Eq. (\ref{Lindbladian}) uses two jump operators, we will need to consider the product of four $\sigma^a_l$ operators, for which there are $4^4=256$ combinations. The coefficients for these operators can be readily derived from the $16\times 16$ combinations of the coefficients $\mu^{s_1,s_2}_j$.

\end{appendix} 


\begin{thebibliography}{32}
\bibitem [{bon()}]{bonddimension}%
  \BibitemOpen
  \href@noop {} {}\bibinfo {howpublished} {In the calculations we use a bond
  dimension $D=100$ and we check that it is converged by simulating also with
  $D=200$.}\BibitemShut {Stop}%
\bibitem [{fn1()}]{fn1}%
  \BibitemOpen
  \href@noop {} {}\bibinfo {howpublished} {For a small system, we have found a
  perfect match when comparing the two-time correlation from the steady state
  obtained from Eq. (\ref{Lindbladian}), when keeping $\gamma=0$ for the
  evolution of the correlation, against the two-time correlation from an exact
  thermal state, for which, also, the evolution of the correlation is only
  computed from Hamiltonian evolution.}\BibitemShut {Stop}%
\end{thebibliography}
\end{document}